\documentclass[aps,pre,superscriptaddress,twocolumn]{revtex4-1}

\usepackage{amsfonts}
\usepackage{amsmath}
\usepackage{multirow}
\usepackage{graphicx}
\usepackage{color}
\usepackage{times}
\usepackage{caption}
\usepackage{subcaption}

\newcommand{\be}{\begin{equation}}
\newcommand{\ee}{\end{equation}}
\newcommand{\bea}{\begin{eqnarray}}
\newcommand{\eea}{\end{eqnarray}}
\newcommand{\av}[1]{\langle #1 \rangle}

\begin{document}

\title{Self-initiated behavioural change and disease resurgence on activity-driven networks}

\author{Nicolò Gozzi}
\affiliation{Networks and Urban Systems Centre, University of Greenwich, London, UK}

\author{Martina Scudeler}
\affiliation{University of Turin, Turin, Italy}

\author{Daniela Paolotti}
\affiliation{ISI Foundation, Turin, Italy}

\author{Andrea Baronchelli}
\affiliation{City, University of London, London, UK}
\affiliation{The Alan Turing Institute, London, UK}

\author{Nicola Perra}
\email[]{n.perra@greenwich.ac.uk}
\affiliation{Networks and Urban Systems Centre, University of Greenwich, London, UK}

\date{\today}

\begin{abstract}
We consider a population that experienced a first wave of infections, interrupted by strong, top-down, governmental restrictions and did not develop a significant immunity to prevent a second wave (i.e. resurgence). As restrictions are lifted, individuals adapt their social behaviour to minimize the risk of infection. We consider two scenarios. In the first, individuals reduce their overall social activity towards the rest of the population. In the second scenario, they maintain a normal social activity within a small community of peers (i.e., social bubble) while reducing social interactions with the rest of the population. In both cases, we consider possible correlations between social activity and behaviour change, reflecting for example the social dimension of certain occupations. We model these scenarios considering a Susceptible-Infected-Recovered epidemic model unfolding on activity-driven networks. Extensive analytical and numerical results show that i) a minority of very active individuals not changing behaviour may nullify the efforts of the large majority of the population, and ii) imperfect social bubbles of normal social activity may be less effective than an overall reduction of social interactions.
\end{abstract}

\maketitle

\section{Introduction}

The spreading of infectious diseases and human behaviour are fundamentally intertwined~\cite{funk,verelst,nine,gross2008adaptive,marceau2010adaptive}. On one side, the unfolding of epidemics might induce people to modify social contacts, habits, and mobility. On the other, such changes might drastically affect the course of the outbreak. \\
\emph{Behavioural change} is a blanket term used to describe a wide range of (re)actions. More in detail, these can be classified into two main categories. The first consists of bottom-up, self-initiated changes implemented by individuals according to their perceived risk and susceptibility as well as to the perceived barriers and benefits linked to each action~\cite{hbm1,hbm2,west2020applying,gozzi2020towards}. These individual decisions vary from social distancing and increased hygiene to the adoption of healthy diets and the use of personal protective equipment such as face masks~\cite{funk,verelst}. The second category, instead, describes top-down, governmental interventions aimed at interrupting chains of infection by banning (or limiting) large gatherings, mobility within and across countries, as well as strict lockdown and cordon sanitarie~\cite{funk,verelst,guzzetta2020impact}.\\
The literature on the subject provides a wealth of theoretical models developed to capture behavioural change and characterise their effects on the disease~\cite{funk,verelst}. These studies differ according to the level of analysis, from single homogeneously mixed populations to individual based contact networks, and according to the mechanisms adopted to model changes in behaviours. Several works tackle the problem by considering variations in individuals' features or in diseases' parameters~\cite{perra2011towards,moinet2018effect,meloni2011modeling,chinazzi2020effect,kraemer2020effect,hebert2020spread}, while others focus on changes in connectivity patterns~\cite{granell2013dynamical,mancastroppa2020active, gross2008adaptive,gross2006epidemic, rizzo2014effect}.  Across the board, such variations are linked to i) disease prevalence and/or ii) individuals' beliefs and (mis)information circulating in the system. The first approach, typically, does not affect the threshold properties of the spreading. In fact, in this case, behavioural change start to be implemented only after the initial growth of the infected population. Nonetheless, prevalence-induced behavioural change can drastically reduce the final disease burden. The second approach, instead, can also affect threshold properties and thus modify the conditions necessary for a macroscopic outbreak even in the case of simple, homogeneously mixed populations~\cite{perra2011towards,wang2016statistical}.  \\
After the first COVID-19 wave, many countries gradually lifted the top-down measures implemented to curb the spreading of the SARS-CoV-2 virus. Such interventions have been largely induced by the local spreading and in particular by the burden to the healthcare systems. When such measures were relaxed, self-initiated behavioural change (nudged by new regulations) became fundamental. Indeed, evidence from serological studies and modelling efforts indicated that the immunity resulting from the first wave was very far from the one required for herd immunity~\cite{Herzog2020.06.08.20125179,Stringhini2020.05.02.20088898,o2020age}. In this context, we present a theoretical framework aimed at investigating the effects of behavioural changes on disease resurgence on time-varying contact networks~\cite{holme2012temporal,holme2015modern,masuda2017temporal}. In particular, we consider the following, unfortunately realistic, scenario. We imagine a population that experienced a first wave of infections which due to strict, top-down measures, was interrupted early. We imagine that interventions are lifted and that people reduce their social interactions (respect to the usual baseline) either because mindful of the risk of propagating the virus (if infected) or because concerned by the risk of infection (if not infected). In doing so, we explore the effects on disease resurgence if such changes are implemented only by a fraction of the population selected i) at random and ii) according to the propensity individuals have to establish social interactions. We adopt activity-driven networks, a class of time-varying networks, to model the temporal interactions between individuals~\cite{perra2012activity,karsai2014time,ubaldi2016asymptotic,ubaldi2017burstiness,alessandretti2017random,nadini2018epidemic}. Although they are a simple approximation of real contact networks, they capture an important property of human interactions: the heterogeneity of human activity. In fact, evidence from a wide range of real datasets capturing human interactions in various contexts suggest that the propensity per unit time of people to establish social connections (i.e. the \textit{activity}), is highly heterogeneous~\cite{perra2012activity,karsai2014time,ubaldi2016asymptotic,ribeiro2013quantifying,tomasello2014role}. Here, we first consider the basic formulation of the model in which, at each time step, active nodes create random connections with others~\cite{perra2012activity}. In these settings, we derive the analytical expression of the epidemic threshold of a Susceptible-Infected-Recovered model~\cite{keeling2011modeling} unfolding on top of the temporal networks as a function of the parameters and mechanisms defining the behavioural changes. Interestingly, we find a closed-form expression for the basic reproductive number $R_0$ that is symmetrical respect to the reduction of activity of susceptible and infected nodes. Since the isolation of large numbers of individuals comes at high societal costs, this underlines the advantages of a configuration with strong isolation of infected. This point becomes even more relevant at the early stage of a possible second wave of infections when the number of infected is relatively small.
Furthermore, we find that in the case of partial adoption assigned in increasing order of activity an almost perfect level of conformity is required to avoid disease resurgence. This highlights that the lack of adoption of a small number of highly socially active individuals may jeopardize large collective efforts.
We then consider a more realistic variation of activity-driven networks able to capture the mesoscopic organisation of real sociograms in tightly connected groups (i.e. communities)~\cite{nadini2018epidemic,fortunato2010community}. In these settings, we rely on numerical simulations to characterise the effects of behavioural changes. In doing so, we consider two different types of adaptive behaviours. The first is a reduction of activity. The second instead is inspired by the concept of social bubbles: nodes keep their social propensity but they direct it towards a small social group. We model this scenario by increasing the probability of interactions within communities. Results show that the presence of communities increases the threshold making it more difficult for a disease to spread. Furthermore, behavioural changes aimed at reducing activity have a much stronger effect on the spreading with respect to those aimed at increasing the cohesiveness of small social groups.\\
The paper is organised as follows. In Section \ref{ADN} we introduce activity-driven networks. In Section \ref{epiADN} we describe the spreading of infectious diseases on this class of time-varying networks. In doing so, we first describe the various mechanisms of behavioural changes induced in the population and then characterize their effects on the spreading of the disease. In Section \ref{conclusion} we present our conclusions.\\

\section{Activity-driven networks}
\label{ADN}

In activity-driven networks, the temporal interactions between $N$ nodes are defined in two steps. The first is the node activation defining the subset of nodes that, at each time step, are active and willing to establish social interactions. The second is the partner selection defining with whom each active node will connect. Nodes' activation is modelled assigning to each node an activity $a$. This quantity, extracted from a distribution $F(a)$, describes the rate at which each node is active per unit time. As mentioned above, observations in several real networks suggest that such distribution is heterogeneous~\cite{perra2012activity,karsai2014time,ubaldi2016asymptotic,ribeiro2013quantifying,tomasello2014role}. For simplicity, in the following we assume $F(a)\sim a^{-\alpha}$ with $\epsilon \le a \le 1$ to avoid divergences. Several mechanisms have been proposed for the second step~\cite{perra2012activity,karsai2014time,ubaldi2016asymptotic,ubaldi2017burstiness,alessandretti2017random,nadini2018epidemic,brett2019spreading}. Here, we consider two: random and community-based partner selection.

\subsection{Random partner selection}

In the basic formulation of the model, each active node creates $m$ random connections with others~\cite{perra2012activity}. Connections are done without recollection of past interactions. Thus, the process is memoryless. In these simple settings, the network's temporal dynamics can be summarised as follows:
\begin{itemize}
\item at each time $t$, the network $G_t$ is initially disconnected;
\item each node is active with probability $a\Delta t$ creating $m$ random connections;
\item each connection is deleted, time incremented to $t+\Delta t$, and the process restarts from the first point.
\end{itemize}
In other words, each connection lasts for a $\Delta t$ duration (without loss of generality here we set $\Delta t=1$) and it is created randomly by active nodes. Thus, at each time step, the network $G_t$ is mostly made up of disconnected stars centered around active nodes. It can be easily shown that the distribution of the number of connections of each node (i.e. the degree) in the aggregated network obtained integrating links over several time-steps follows the distribution of the activity~\cite{perra2012activity,starnini2013topological}. Hence, heterogeneous activity patterns induce the formation of hubs which are highly active nodes engaging over and over in social interactions. However, since links are created at random, the distribution of links' weights in the time-integrated network is homogeneous and thus very far from observations in real networks~\cite{onnela2007structure,karsai2011small,miritello2011dynamical}. In summary, this version of the model captures some important features of real systems and it allows for analytical analyses of dynamical processes unfolding on its structure at comparable time-scales~\cite{perra2012activity,perra2012random,liu2014controlling,davis2020phase,starnini2014temporal,rizzo2014effect,zino2016continuous,zino2017analytical}, but at the same time it is a rough approximation of real social networks.

\subsection{Community-based partner selection}

This second approach considers a much more realistic partner selection mechanism. In fact, social networks are organised in tightly connected groups (i.e. communities), which emerge and evolve in time~\cite{fortunato2010community}. As a result, the vast majority of connections takes place within such circles of friends rather than across them~\cite{karsai2011small,onnela2007structure}. To capture this fundamental aspect of social interactions, each node is assigned to a particular community $c$. The size $s$ of each community is extracted from a distribution $G(s)$~\cite{nadini2018epidemic}. In these settings, the dynamics of the network follows these steps:
\begin{itemize}
\item at each time $t$, the network $G_t$ is initially disconnected;
\item each node is active with probability $a\Delta t$ creating $m$ connections;
\item with probability $\eta$ each connection is done selecting at random one of the nodes in the same community and with probability $1-\eta$  selecting at random in any other community;
\item each connection is deleted, time incremented to $t+\Delta t$ and the process restarts from the first point.
\end{itemize}

As done above, without loss of generality we set $\Delta t=1$. The parameter $\eta$ regulates the modularity of the emerging network. For $\eta=0$ (and in case of community sizes $s \ll N$) the network unfolds very similarly to the first model. Instead for $\eta=1$ the network will be formed by completely disconnected communities.

\section{Behavioural changes induced by disease spreading on activity-driven networks}
\label{epiADN}

We study the spreading of an infectious disease unfolding at a comparable time-scale with respect to the evolution of connections in the contact network. We consider the prototypical Susceptible-Infected-Recovered epidemic model~\cite{keeling2011modeling,pastor2015epidemic}. Thus, each node can be found in one of three compartments: healthy and susceptible individuals are in the compartment $S$, infectious in $I$, and recovered in $R$. The disease propagates via the connections between susceptible and infectious nodes with a probability of infection, per contact, $\lambda$. Infected nodes recover spontaneously with probability $\mu$. In the case of an outbreak without residual immunity (i.e. $R(t=0)=0$), the epidemic threshold in a memoryless activity-driven network can be obtained, using a mean-field approach, studying the evolution of the number of infected nodes with activity $a$. In particular, this threshold reads~\cite{perra2012activity}:
\be
\label{activity_threshold}
\frac{\lambda}{\mu}> \frac{1}{m}\frac{1}{\av{a}+\sqrt{\av{a^2}}}
\ee 
which implies that the disease will be able to spread only if $R_0= m\frac{\lambda}{\mu}\left (\av{a}+\sqrt{\av{a^2}} \right )>1$. In this last expression, we introduced the basic reproductive number $R_{0}$ as the number of secondary infections generated by an index case in a fully susceptible population~\cite{keeling2011modeling}. Interestingly, the threshold of a SIR (for a SIS the threshold is the same) model is driven by the first and second moment of the activity distribution rather than the time-aggregated properties of the graph. \\
In the case of activity-driven networks with communities, we do not have a closed expression for the threshold. However, it is interesting to notice how the presence of communities, and more in general the repetition of a small subset of ties~\cite{sun2015contrasting,tizzani2018epidemic}, affects very differently SIR and SIS models~\cite{nadini2018epidemic}. In fact, while in the case of permanent immunity (SIR) the repetition of contacts within communities hampers the spreading, it helps the diffusion of diseases able to reach an endemic state (SIS). In other words, modularity pushes the threshold of SIR models to higher values while facilitates the spreading of SIS models, pushing the threshold to lower values.

As mentioned in the introduction, in this work we investigate the following scenario. A highly infectious disease spreads in the network, but its course is halted by strict top-down interventions. In this first wave, the large majority of individuals has not been affected, thus the system is far from herd immunity (i.e. $R \ll N$). As the measures are lifted, individuals (maybe nudged by laws and regulations) implement behavioural changes with the aim of protecting themselves and the others and thus avoid - or mitigate - a second wave.\\
As a first step, we model the implementation of behavioural changes reducing the activity of susceptibles by a factor $\gamma$ and the activity of infected by a factor $\psi$. This model has been presented first by Rizzo et al~\cite{rizzo2014effect}. As a second step, we extend this approach by studying the effect of adoption rates. We consider that such changes in behaviour are implemented only by a fraction of nodes selected either at random or as a function of the activity. In fact, the adoption of behavioural changes can be more costly for very active people (maybe due to their occupation) and thus less likely to be applied~\cite{hbm1,hbm2}. As a third step, we further extend the literature exploring the interplay between behavioural changes and the modularity of the network. This corresponds to the fact that individuals might keep the same activity but cut connections with people outside their close-knit social circles. Also in this case, we study the role of adoption considering only a fraction of nodes engaging in any form of behavioural change.

\subsection{Perfect adoption}

Following the order described above, we first consider a scenario in which, as a way to reduce the risk of infection, susceptible individuals reduce their activity of a factor $\gamma$ and infected by a factor $\psi$. Assuming that nodes of the same activity are statistically equivalent, we can write the evolution of the number of infected node of activity class $a$ as: 
\bea
d_t I_a &=& -\mu I_a +m \lambda (N_a-I_a-R_a) a \gamma \int da' \frac{I_{a'}}{N} + \\ \nonumber
&+& m \lambda (N_a-I_a-R_a)\psi  \int da' \frac{I_{a'}a'}{N}
\eea
where $N_a=S_a+I_a+R_a$ holds for all activity classes. The first term in the right hand side describes recovery process. The second captures susceptible nodes that become active and select as partner an infected individual in any other activity class. The third term, instead, describes susceptible nodes that are selected by active and infected nodes in any activity class. In the early stages of the possible second wave, we assume $I_a\ll S_a$ and $R_a \ll S_a$ thus $N_a \sim S_a$. 

In other words, the first wave was stopped well before the disease was able to affect a large fraction of the population. In these settings we drop all the second order terms obtaining: 
\be
\label{eqia}
d_t I_a = -\mu I_a +m \lambda N_a a \gamma \int da' \frac{I_{a'}}{N}
+ m \lambda N_a\psi  \int da' \frac{I_{a'}a'}{N}.
\ee
By integrating both sides over all activity classes we have: 
\be
\label{eqi}
d_t I = -\mu I +m \lambda \gamma \av{a} I
+ m \lambda \psi  \Theta
\ee
where $\Theta=\int da I_{a}a$ and $\av{a^n}=\int da F(a)a^n$. To characterise the evolution of $I(t)$ we then need to derive an equation for $\Theta$. In particular, multiplying both sides of Eq.~\ref{eqia} by $a$ and integrating across all activities:
\be
\label{eqtheta}
d_t \Theta = -\mu \Theta +m \lambda \gamma \av{a^2} I
+ m \lambda \psi \av{a}  \Theta.
\ee

The epidemic threshold can be obtained studying the stability of the system of differential equations defined by Eq.~\ref{eqi} and Eq.~\ref{eqtheta}. Indeed, the disease will be able to spread only if the largest eigenvalue of the Jacobian matrix $J$ of the system is larger than zero. Here, $J$ can be written as:
\begin{equation*}
J = 
\begin{pmatrix}
-\mu + m \lambda \gamma \av{a}& m\lambda \psi   \\
m \lambda \gamma \av{a^2} & -\mu + m \lambda \psi \av{a}
\end{pmatrix}
\end{equation*}
And the threshold, as reported in Ref.~\cite{rizzo2014effect}, is:
\be
\label{first_thre}
\frac{\lambda}{\mu}> \frac{2}{m}\frac{1}{\av{a}(\gamma+\psi)
+\sqrt{(\gamma-\psi)^2\av{a}^2+4\gamma\psi\av{a^2}}}.
\ee
This condition implies $R_0=(\lambda/\mu) \mathcal{T}^{-1}>1$, where $\mathcal{T}$ is the right hand side Eq.~\ref{first_thre}. Notably, when $\gamma=\psi=1$, it reduces to the threshold of activity-driven networks without behavioural changes. However, it is interesting to notice how these two reductions of activity rates do not imply a simple rescaling of the threshold. Indeed, they introduce non-linear terms in the expression. Furthermore, the threshold is symmetric in $\gamma$ and $\psi$ meaning that the reduction in activity of susceptibles ($\gamma$) and of infected ($\psi$) can be switched without implying any change in the threshold. It is important to notice, however, that in the early stages of the outbreak the number of susceptibles is much larger than the number of infected. Thus, since reductions of activity come with high social costs, combinations of parameters with lower values of $\psi$ rather than $\gamma$ are desirable.  \\
In Fig.~\ref{fig1}-a we show the basic reproductive number $R_0$ obtained from Eq.~\ref{first_thre} as a function of $\gamma$ and $\psi$. We consider a scenario in which, without behavioural changes (i.e. for $\gamma=\psi=1$), we have that $R_0=3$. Hence, the disease is supercritical and able to spread in the population. Introducing increasingly stronger behavioural changes implemented by susceptible and infected (i.e. as both $\gamma$ and $\psi$ decrease), $R_0$ becomes progressively smaller. The red line in figure splits the parameters space in two: to its left, we find the combinations of $\gamma$ and $\psi$ such that $R_{0} < 1$, while to its right we have those combinations that are not strong enough to push $R_{0}$ below $1$ and thus prevent the onset of a second wave. From the plot, we see that when $\gamma=1$ (i.e. susceptible nodes do not implement any behavioural change), infected nodes need to reduce their activity to less than $10\%$ of the original in order to stop the resurgence of the disease.

\subsection{Non-perfect adoption}

\begin{figure}
  \centering
    \begin{subfigure}[b]{0.49\linewidth}
    \centering
    \includegraphics[width=\textwidth]{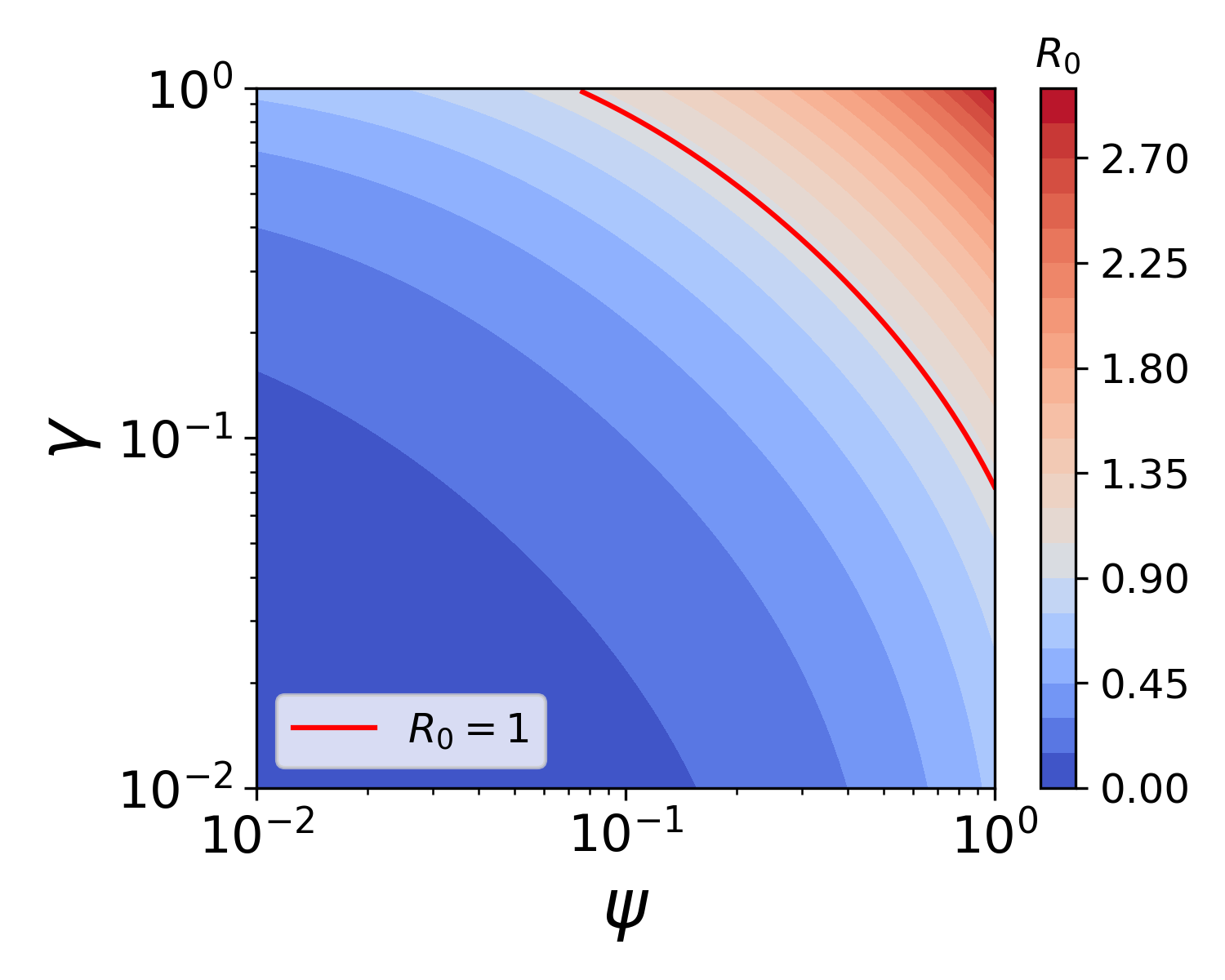}
    \caption{}
    \label{1}
  \end{subfigure}
  \begin{subfigure}[b]{0.49\linewidth}
    \centering
    \includegraphics[width=\textwidth]{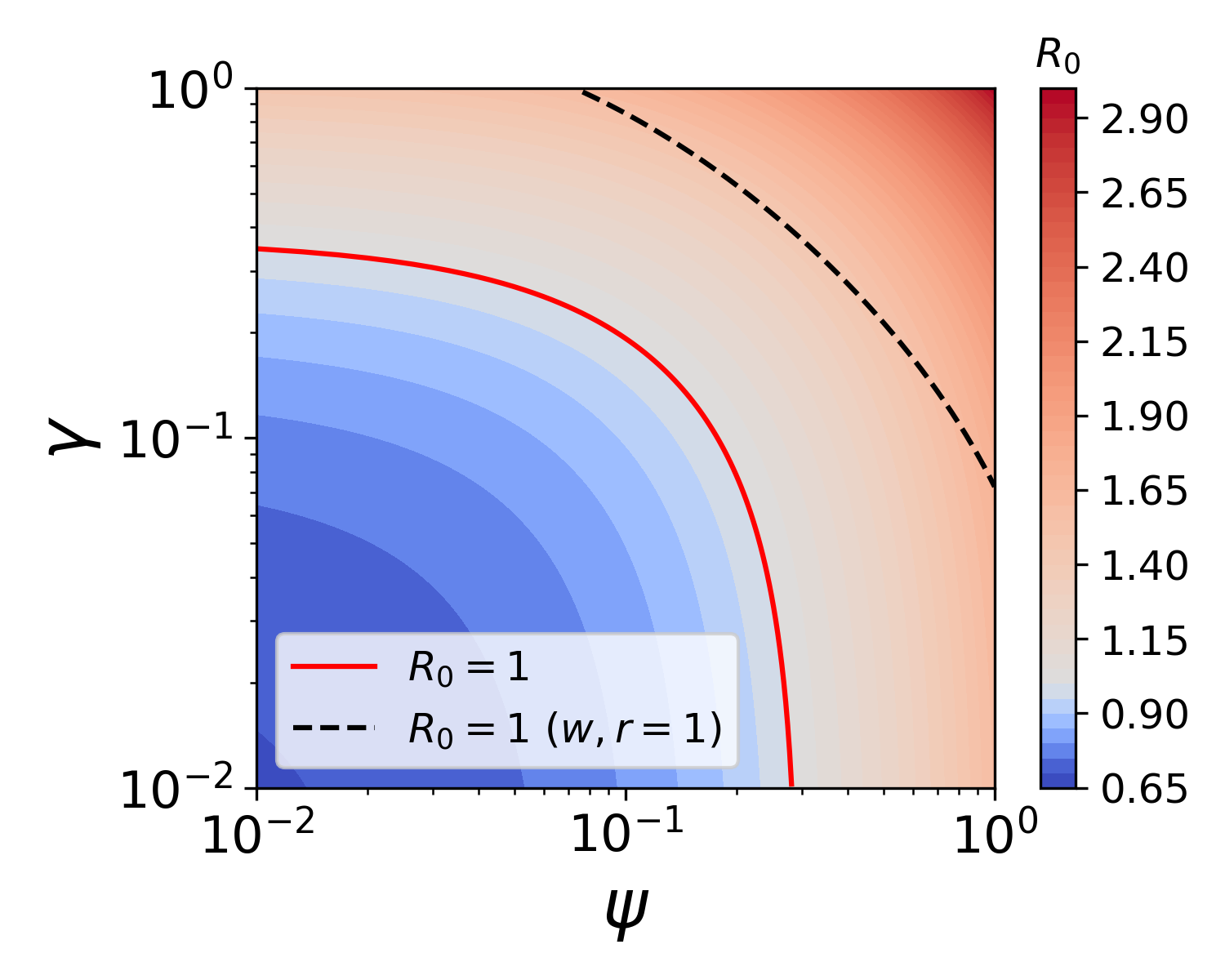}
    \caption{}
    \label{2a}
  \end{subfigure}
  \begin{subfigure}[b]{0.49\linewidth}
    \centering
    \includegraphics[width=\textwidth]{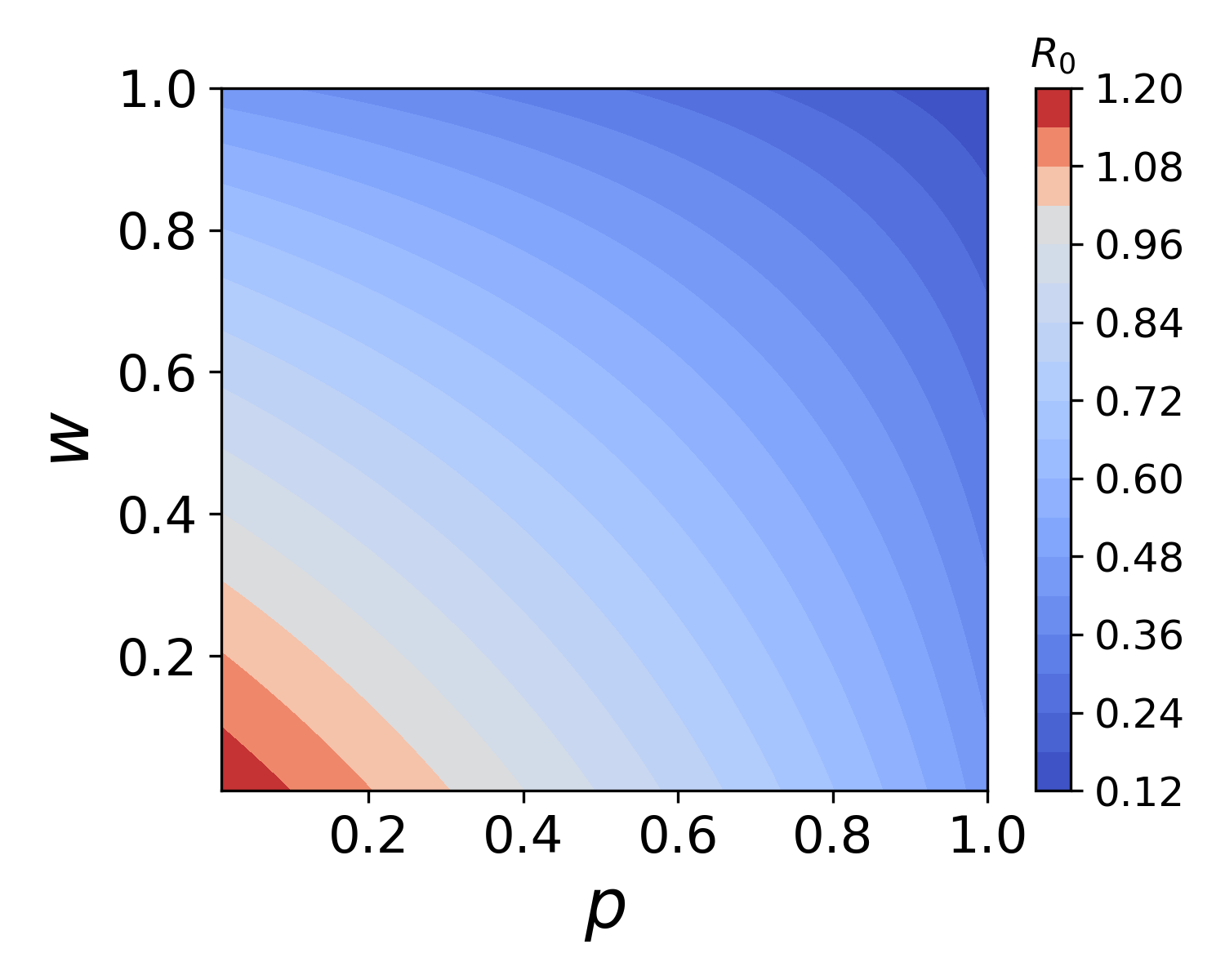}
    \caption{}
    \label{2b}
  \end{subfigure}
    \begin{subfigure}[b]{0.49\linewidth}
    \centering
    \includegraphics[width=\textwidth]{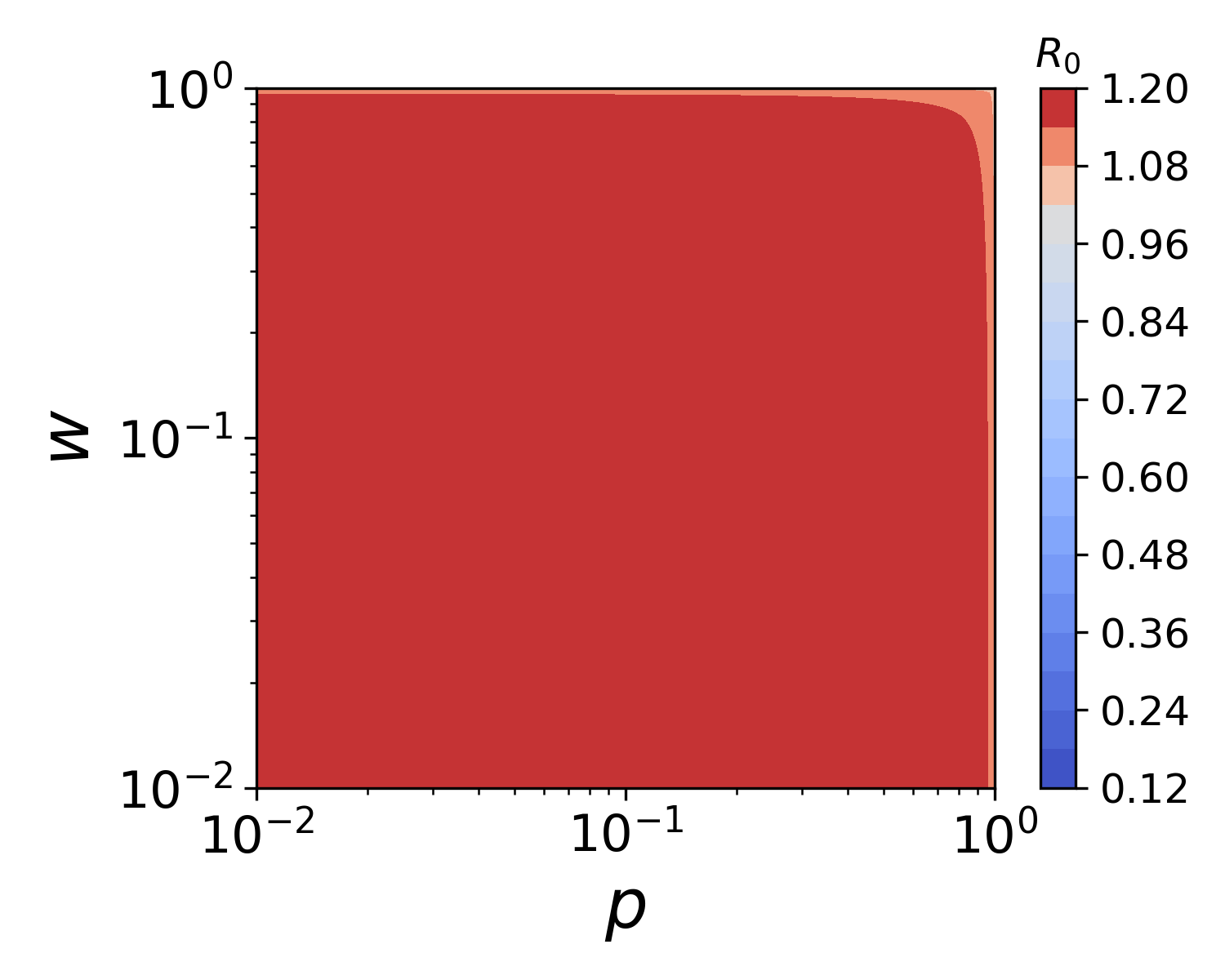}
    \caption{}
    \label{4}
  \end{subfigure}
  \caption{We show the analytical value of the basic reproductive number $R_0$ in different scenarios. a) $R_0$ as a function of $\gamma$ and $\psi$ as obtained in Eq.~\ref{first_thre} in the case of perfect adoption to behavioural change. b) $R_0$ as a function of $\gamma$ and $\psi$ as obtained in Eq.~\ref{thre2} in the case of non-perfect adoption independent of nodes' activity. We set $p = 0.75$ and $w = 0.8$. c) $R_0$ as a function of $p$ and $w$ as obtained in Eq.~\ref{thre2} in the case of non-perfect adoption independent of nodes' activity. We set $\gamma = 0.1$ and $\psi = 0.1$ d) $R_0$ as a function of $p$ and $w$ as obtained in Eq.~\ref{thre3} in the case of adoption dependent from nodes' activity. We set $\gamma = 0.1$ and $\psi = 0.1$, and use the same $\lambda$ of panel c). In all figures, we set $\epsilon = 10^{-3}$, $m = 2$, $\alpha = 2.1$, $ \mu = 10^{-2}$. In panels a) and b) we set $R_0=3$ and in panels c) and d) $R_0=1.2$ when $\gamma=1$, $\psi=1$ (i.e. without behavioural changes).}
  \label{fig1}
\end{figure}

So far we have assumed that all nodes modify their behaviours in response to the outbreak. However, it is more realistic to assume that only a fraction of the population is willing - or able - to reduce the activity. \\
For this reason, we first consider that the propensity to implement behavioural changes is independent of node's features and only function of its status. Thus, we can assume that, across all activity classes, a fraction $p$ of susceptibles and a fraction $w$ of infected will reduce the activity. In these settings, it can be easily shown that the system of differential equations describing the early-stage dynamics of the disease is equal to the one derived before provided these two transformations: $\gamma \rightarrow \gamma_p=1-p(1-\gamma)$ and $\psi \rightarrow \psi_w=1-w(1-\psi)$. Hence, the threshold can be rewritten as:
\be
\label{thre2}
\frac{\lambda}{\mu}> \frac{2}{m}\frac{1}{\av{a}(\gamma_p+\psi_w) 
+\sqrt{(\gamma_p-\psi_w)^2\av{a}^2+4\gamma_p\psi_w\av{a^2}}}.
\ee
In other words, the case of random adoption across classes does not alter the dynamics but simply rescales the reduction parameters. Not surprisingly, obtaining the same increase of the threshold requires an extra reduction of activity of the fraction of the population that complies. We can see these effects in Fig.~\ref{fig1}-b where we show $R_0$ (calculated from Eq.~\ref{thre2}) as a function of $\gamma$ and $\psi$ for fixed values of $p$ and $w$. As expected, in the case of random adoption the threshold is shifted towards lower values of $\gamma$ and $\psi$ with respect to the previous case (grey dashed line). This implies that larger reductions of activity are required in order to stop the disease from spreading. Furthermore, since $p$ and $w$ have different values and therefore weighs differently the reduction parameters, the $R_0$ is not symmetrical in $\gamma$ and $\psi$ (but in $\gamma_p$ and $\psi_w$). 
In Fig.~\ref{fig1}-c, instead, we show $R_0$ as a function of $p$ and $w$ for fixed values of $\gamma$ and $\psi$. We observe that $p$ and $w$ interpolate between two opposite regimes. By reducing their values, the adoption of behavioural changes becomes increasingly less significant. Consequently, $R_0$ assumes values closer to those described in Eq.~\ref{activity_threshold} which captures a system without behavioural changes. Conversely, by increasing the values of $p$ and $w$, the values of $R_0$ become progressively closer to Eq.~\ref{first_thre} which describes a system with $100\%$ adoption of behavioural changes. \\

\begin{figure}
  \centering
    \begin{subfigure}[b]{0.49\linewidth}
    \centering
    \includegraphics[width=\textwidth]{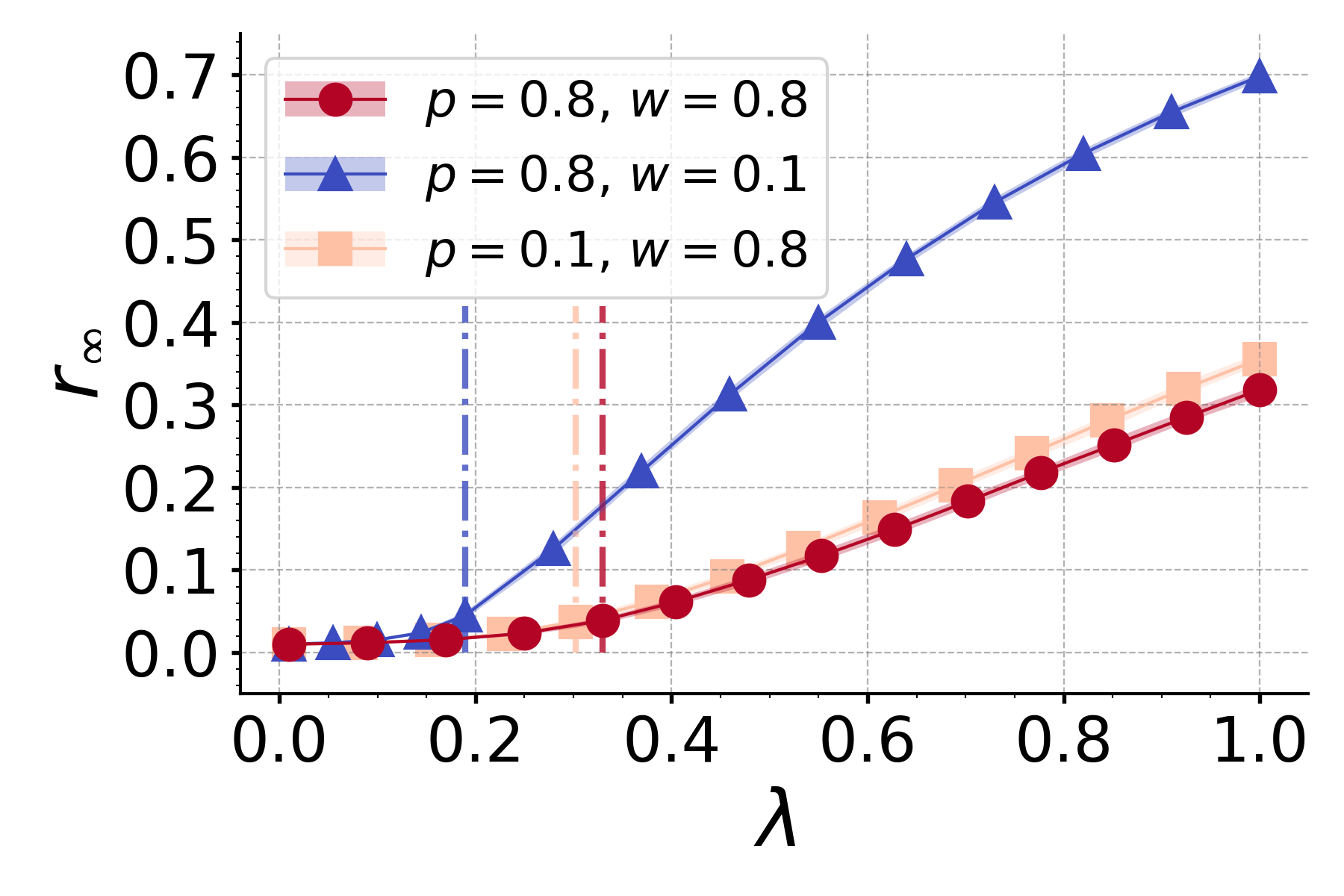}
    \caption{}
    \label{pr}
  \end{subfigure}
  \begin{subfigure}[b]{0.49\linewidth}
    \centering
    \includegraphics[width=\textwidth]{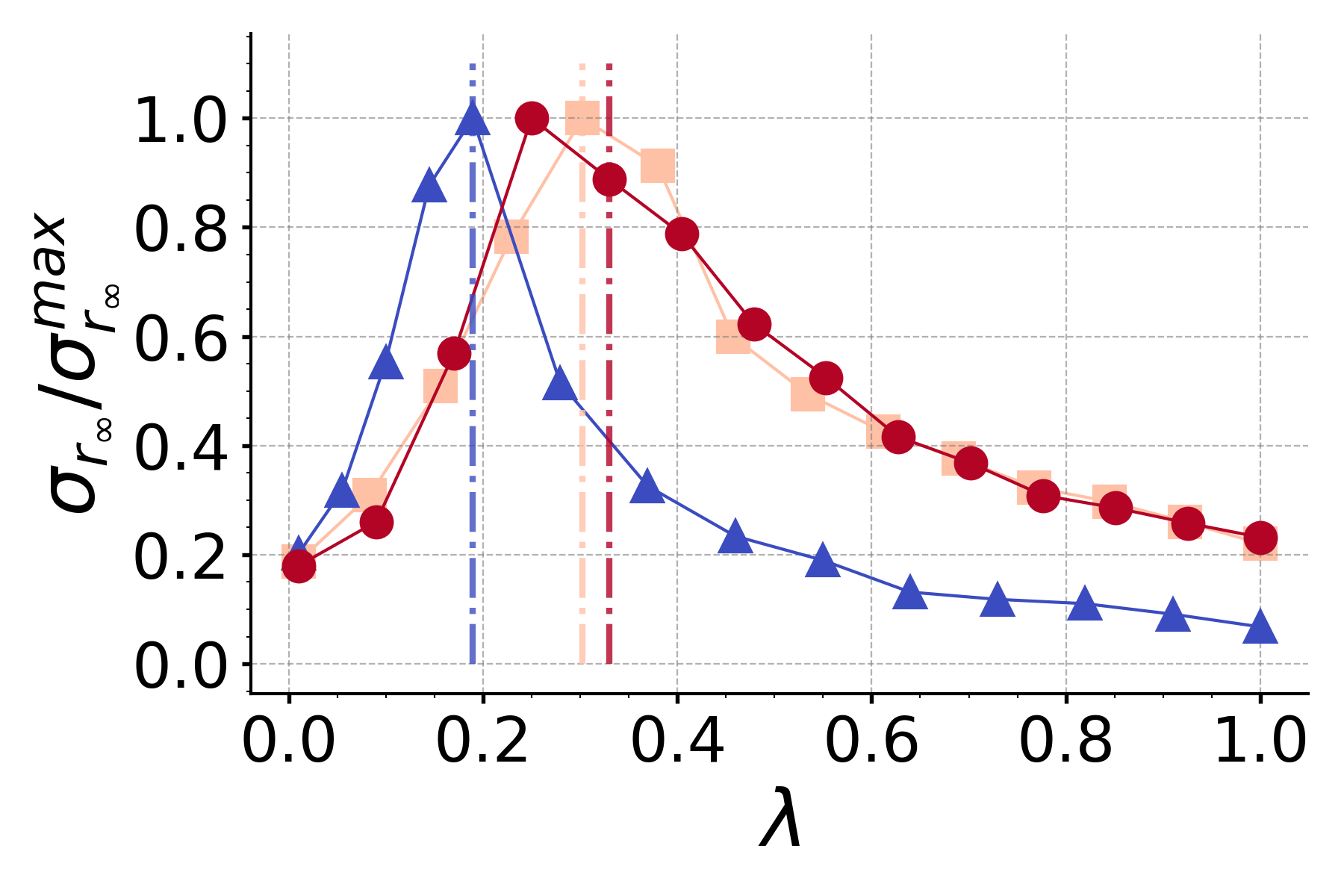}
    \caption{}
    \label{pr_sigma}
  \end{subfigure}
  \begin{subfigure}[b]{0.49\linewidth}
    \centering
    \includegraphics[width=\textwidth]{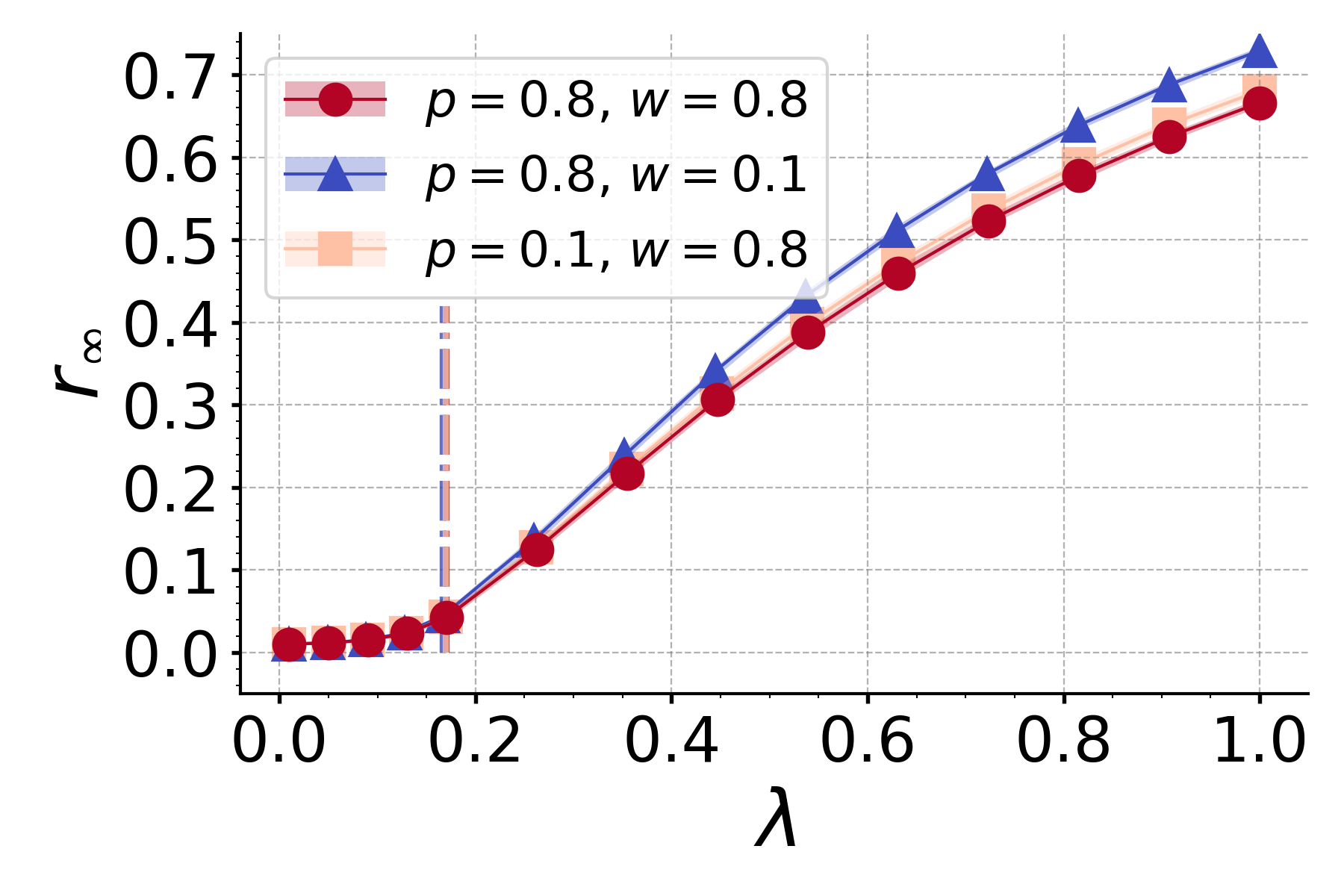}
    \caption{}
    \label{apar}
  \end{subfigure}
    \begin{subfigure}[b]{0.49\linewidth}
    \centering
    \includegraphics[width=\textwidth]{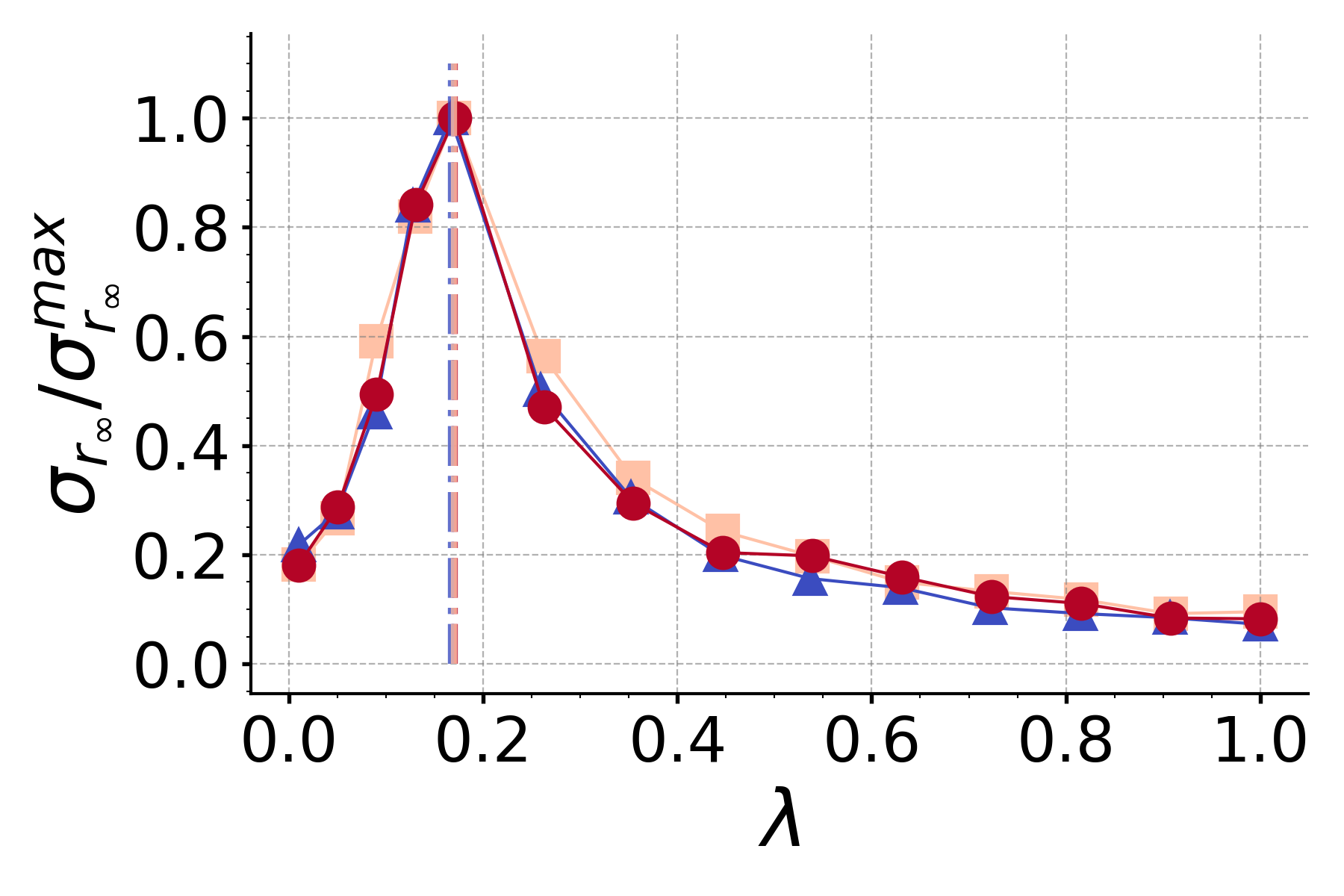}
    \caption{}
    \label{apar_sigma}
  \end{subfigure}
  \caption{In panel a) we display the final epidemic size ($r_\infty=R_\infty / N$) with $95\%$ confidence intervals and in panel b) the normalized relative variance $\sigma_{r_\infty} / \sigma_{r_{\infty}}^{max}$ for different values of $\lambda$ in the case of non-perfect random adoption. Vertical dashed lines indicates the analytical threshold derived from Eq.~\ref{thre2} for the different values of $p$ and $w$ considered. In panel c) and d) we repeat the analysis in the case of non-perfect adoption dependent from nodes’ activity for different values of $p$ and $w$. Results are obtained from $10^{2}$ stochastic simulations for each point and with the following model parameters: $\mu=10^{-2}$, $N=10^{6}$, $m=2$, $\epsilon=10^{-3}$, $\alpha=2.1$, $i_0=I_0/N = 0.01$ (initial fraction of infected seeds), $\gamma=0.8$, $\psi=0.1$.}
  \label{fig2}
\end{figure}

We test the analytical solution derived in Eq.~\ref{thre2} by means of numerical simulations. In particular, we consider a case in which infected individuals reduce more their activity (due to their illness status, for example) with respect to the susceptibles by setting $\gamma=0.8$, and $\psi=0.1$. In Fig.~\ref{fig2}-a, we plot $r_\infty=R_\infty / N$ as a function of $\lambda$, for different values of $p$ and $w$. Reasonably, the epidemic size grows with the infectiousness of the pathogen and it decreases when a larger fraction of individuals implement behavioural changes. As expected from the theory, the final epidemic size obtained with $80\%$ of susceptibles and infected implementing behavioural changes is very similar to the one obtained with only $10\%$ of susceptibles and $80\%$ of infected engaging in such changes. Taking into account that at the beginning of the possible second wave the number of infected is much smaller than the number of susceptible, and given the high socio-economic cost of isolating individuals, a setting in which infected reduce more their social interactions is clearly more desirable. In Fig.~\ref{fig2}-b, instead, we represent, for different combinations of parameters, the relative variance $\sigma_{r_\infty}$ of the final epidemic size as a function of $\lambda$~\cite{moinet2018effect}. This is defined as $\sigma_{r_\infty} = \sqrt{\langle r_\infty^2 \rangle - \langle r_\infty \rangle ^2} / {\langle r_\infty\rangle}$. Because of the critical behaviour of the epidemic process we are considering, the maximum $\sigma_{r_{\infty}}$ is reached at the threshold. As expected, we observe that, for the different combinations of parameters considered, the normalized relative variance peaks around the theoretical threshold values, providing a numerical validation of the analytical expression found previously.

\subsection{Non-perfect adoption depending on activity}

Moving to even more realistic scenarios, we now consider the case in which the propensity of implementing behavioural changes is linked to the activity. We can imagine that, due to the higher barriers and costs associated with the change in behaviour, nodes with larger activity will be less prone to such changes. A natural example are individuals that due to the nature of their job cannot easily reduce the number of their interactions. For simplicity, we consider that only susceptible nodes with activity lower or equal to $a_p$ and infected nodes with activity lower or equal than $a_w$ will implement behavioural change. In other words, nodes with activity larger than a given threshold will not change their behaviour. Therefore, the maximum fraction $x$ of individuals changing behaviour is linked to the activity cut-off value $a_{x}$ as follows:
\be
\label{xfa}
x=\int_{\epsilon}^{a_x}F(a)da.
\ee
In general, since the health status of each node affects the type of behavioural change, the fraction of, say, infected nodes adopting behavioural change is smaller that the respective $x$. \\
In these settings, the equation regulating the change of number of infected in a specific activity class $a$ at early stages of the spreading can be written:
\bea
\label{neweqia}
d_t I_a &=& -\mu I_a +m \lambda N_a a \left [ 1-\delta[a_p-a](1-\gamma) \right ] \int da' \frac{I_{a'}}{N} +  \nonumber \\
&+& m \lambda N_a  \int da' \left [  1- \delta[a_w-a'](1-\psi) \right ]\frac{I_{a'}a'}{N},
\eea
where $\delta[x-y]$ is a heaviside step function equal to one for $x\ge y$ and zero otherwise. Integrating over all activities and introducing $\Xi= \int da' \left [  1- \delta[a_w-a'](1-\psi) \right ] I_{a'}a'$ we get:
\be
d_t I = -\mu I +m \lambda  \left [ \av{a}+ \av{a}_p(1-\gamma) \right ] I + m \lambda \Xi,
\ee
where we defined $\av{a}_p=\int_{\epsilon}^{a_p}F(a)ada$. In order to understand the early dynamics we need to get an equation for $\Xi$. To this end, we can multiply both terms of Eq.~\ref{neweqia} for 
$\left [  1- \delta[a_w-a](1-\psi)\right ]a$ and integrating across all activities we obtain: 
\bea
d_t \Xi &=& -\mu \Xi +m \lambda \mathcal{F}[\av{a^2},\gamma,\psi,a_p,a_w] I +  \nonumber \\
&+& m \lambda \left [ \av{a} - \av{a}_w(1-\psi)\right ]\Xi,
\eea
where we defined:
\bea
\label{F_modulation}
\mathcal{F}[\av{a^2},\gamma,\psi,a_p,a_w]&=& \av{a^2}-\av{a^2}_p(1-\gamma)-\av{a^2}_w(1-\psi) \nonumber \\
&+& \av{a^2}_{min_{a_p,a_w}} (1-\gamma)(1-\psi)
\eea
which is a modulation of the second moment of the activity distribution. From this stand point we can write the Jacobian matrix of the system of differential equations. To simplify further the notation, we define:

\begin{align}
    \label{alpha_p}
    &\alpha_p = \langle a \rangle - \langle a \rangle_p(1-\gamma) \\  
    \label{alpha_w}&\alpha_w = \langle a \rangle - \langle a \rangle_w(1-\psi) 
\end{align}

Therefore, the threshold behaviour is encoded in the system of differential equations:

\begin{align}
    d_t I &=  (-\mu + \lambda m \alpha_p) I + \lambda m \Xi \\ \nonumber \\
    d_t \Xi &=  + \lambda m \mathcal{F} I + ( -\mu + \lambda m \alpha_w ) \Xi
\end{align}

The disease will be able to grow only if the largest eigenvalue of the Jacobian matrix of this system is larger than zero. The Jacobian matrix can be written as:
 \begin{equation}
     J = \begin{pmatrix}
            -\mu + \lambda m \alpha_p & \lambda m\\
            \lambda m \mathcal{F} & -\mu + \lambda m \alpha_w
        \end{pmatrix}
 \end{equation} 

The eigenvalues $k_{1,2}$ can be found solving the characteristic equation:

\begin{align}
    &k^2 + k \left(2\mu - \lambda m (\alpha_p + \alpha_w) \right) + \lambda^2 m^2 \left(\alpha_p \alpha_w - \mathcal{F} \right)  \nonumber\\ & - \mu \lambda m \left(\alpha_p+ \alpha_w \right) + \mu ^2 = 0
\end{align}

leading to:

\begin{equation}
    k_{1,2} = \frac{1}{2} \left[ -2\mu + \lambda m \left(\alpha_p + \alpha_w \right) \pm \lambda m \sqrt{\left(\alpha_p - \alpha_w\right)^2 + 4 \mathcal{F}}\right]
\end{equation}

The threshold condition can be written as:

\begin{equation}
    \frac{\lambda}{\mu} > \frac{2}{m \left[ \left(\alpha_p + \alpha_w \right) + \sqrt{\left(\alpha_p - \alpha_w \right)^2 + 4 \mathcal{F}}\right] } 
\end{equation} \\

Substituting the values of $\alpha_p, \alpha_w,  \mathcal{F}$ with those from equation \ref{F_modulation}, \ref{alpha_p}, \ref{alpha_w}, we obtain the threshold as a function of the activity moments:

\begin{equation}
    \frac{\lambda}{\mu} > \frac{2}{m \left[ 2 \langle a \rangle - \left(1 - \gamma \right) \langle a \rangle_p - \left(1 - \psi \right) \langle a \rangle_w + \sqrt{\Delta}\right] }
    \label{thre3}
\end{equation} \\

where 

\bea
    \Delta &=& 4 \langle a^2 \rangle + (1-\gamma) [\langle a \rangle_p^2 (1-\gamma) - 4\langle a^2 \rangle_p]  \nonumber\\ &+& (1-\psi) [\langle a \rangle_w^2 (1-\psi) - 4\langle a^2 \rangle_w] \nonumber\\ &+& 2 (1 - \gamma)(1 - \psi)[2 \langle a^2 \rangle_{min(a_p,a_w)} - \langle a \rangle_w \langle a \rangle_p]
\eea \\

We can verify that, if $a_p = a_w = 1$, we find again the result of Eq.~\ref{thre2}. In Fig.~\ref{fig1}-d we compare the effects of different values of $p$ and $w$ on $R_0$ in the case of random and activity-based adoption using Eq.~\ref{xfa} to compute the correspondent values of $a_p$ and $a_w$. Very differently from the case of a random fraction of adopting nodes (Fig.~\ref{fig1}-c), an almost perfect conformity to the behavioural measures is needed to halt the spreading and push $R_0$ below $1$. Indeed, even if the majority of individuals reduces the activity, the interactions of the most activity nodes are sufficient for a disease to spread. This is line with the study of immunization strategies in activity-driven networks~\cite{liu2014controlling}. In fact, the strategic immunization of few central, most active nodes has been shown sufficient to halt the spreading. This implies that these nodes are indeed key to sustain the unfolding on the virus. Hence, the highly heterogeneous activity of people plays a fundamental role in this phenomenon. In Fig.~\ref{fig2}-c-d we test the analytical solution by means of numerical simulations. We plot $r_\infty$ and $\sigma_{r_\infty}/\sigma_{r_\infty}^{max}$ as functions of $\lambda$, for different values of $p$ and $w$. Interestingly, we observe that large differences in $p$ and $w$ result in very similar attack rate profiles and thresholds. This confirm how the spreading is controlled by the most active nodes that are not compliant and how the efforts of large majority of the population might be vane if a minority of highly active node does not change behaviour. Also, Fig.~\ref{fig2}-d confirms the validity of the theoretical analysis. Indeed, in all scenarios considered the $\sigma_{r_\infty}/\sigma_{r_\infty}^{max}$ estimated in the simulations are well peaked around the analytical predictions.

\subsection{The role of communities}

Moving forward we switch to the second, more realistic, time-varying network model where the link creation mechanism is function of nodes' membership to tightly connected groups, i.e. communities. The analysis that follows is based only on numerical simulations, indeed as mentioned above we do not have a closed-form expression for the threshold of activity-driven networks with communities even in the absence of behavioural change~\cite{nadini2018epidemic}. The presence of a modular structure allows us to study two different types of changes in behaviours. The first is analogous to what we presented above: as a way to protect themselves and the others, nodes reduce their social propensity. We label this as \emph{activity reduction} (AR). The second instead takes inspiration from the idea of \emph{social bubbles}~\cite{leng2020effectiveness}. Indeed, individuals might keep the same social propensity but direct it only towards a limited group of people. To this end, we hypothesize that nodes keep the same activity but increment the share of intra-community connections leading to an increase in the modularity of the emergent social network. We label this as \emph{modularity increase} (MI).\\

\subsubsection{Activity reduction}

As done above, in this scenario we imagine that individuals change behaviours by reducing their activity by a factor $\gamma$ if susceptible and by a factor $\psi$ if infected. Furthermore, we consider the case of non-perfect adoption and compare random with activity-based adoption. In Fig.~\ref{fig3}-a we show the behaviour of the epidemic size as a function of $\lambda$. We set $\gamma=0.8$, $\psi=0.1$ and $p=w=0.6$.  We consider a modularity $\eta=0.6$, and for simplicity we set the size of each community to be the same ($s=10$). Thus $60\%$ of links are created within communities and each one is made up of ten nodes. Few observations are in order. First, as seen before, random adoption is characterized by a larger threshold and a smaller final epidemic size with respect to the other case. Therefore, when the most active nodes are not adapting their behaviour the system is more fragile to the spreading of a virus. Second, we plot as dashed lines the analytical thresholds computed in absence of communities from Eq.~\ref{thre3}. In both cases, the presence of tightly connected groups of nodes increases the threshold. This result is in line with past research showing that high values of modularity slow down the spreading of SIR (as well as SI) models~\cite{nadini2018epidemic,onnela2007structure,karsai2011small,salathe2010dynamics,stegehuis2016epidemic,scarpino2016effect}.

\subsubsection{Modularity increase}

\begin{figure}[ht!]
  \centering
    \begin{subfigure}[b]{\linewidth}
    \centering
    \includegraphics[width=\textwidth]{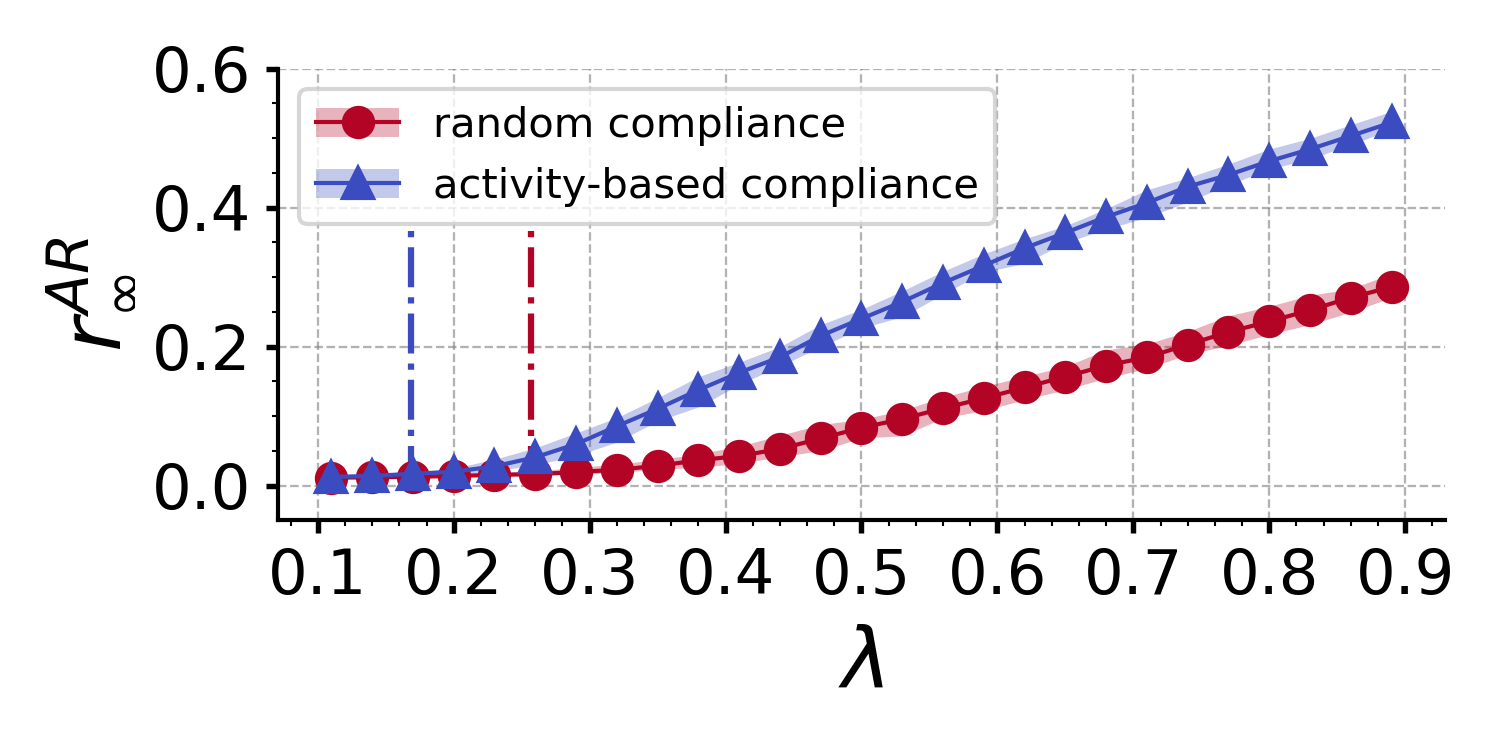}
    \caption{}
    \label{comm_act}
  \end{subfigure}
  \begin{subfigure}[b]{\linewidth}
    \centering
    \includegraphics[width=\textwidth]{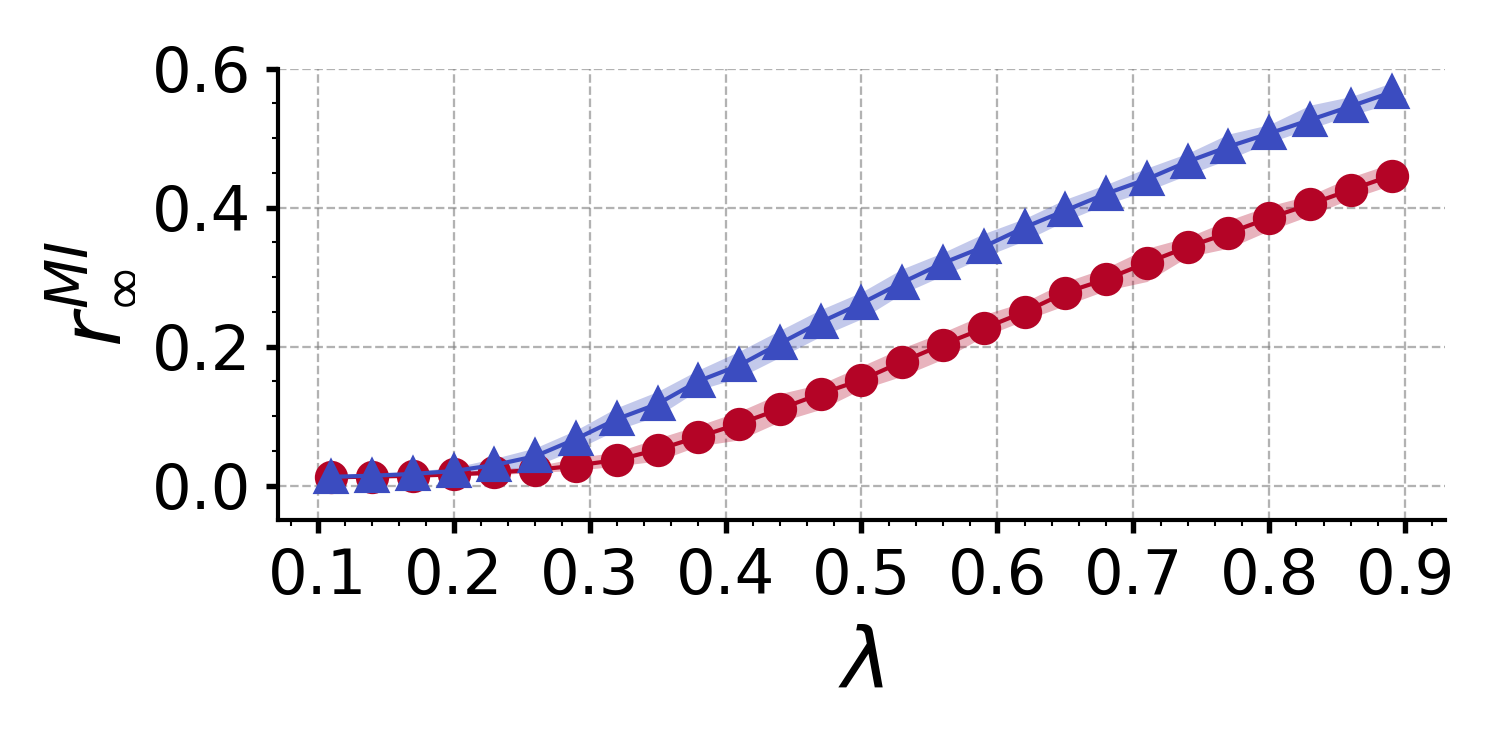}
    \caption{}
    \label{comm_mod}
  \end{subfigure}
  \begin{subfigure}[b]{\linewidth}
    \centering
    \includegraphics[width=\textwidth]{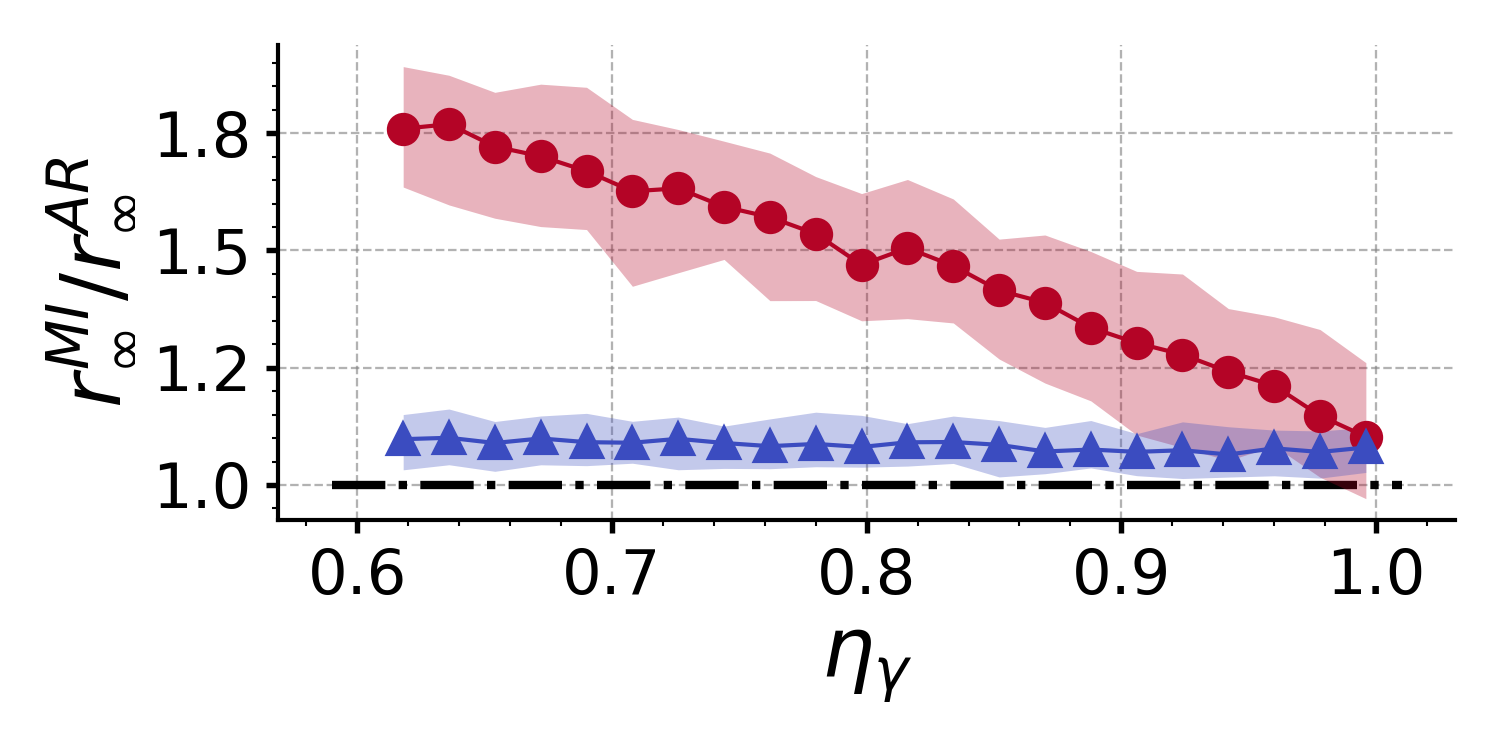}
    \caption{}
    \label{ratio}
  \end{subfigure}
  \caption{a) Epidemic size as a function of $\lambda$ in presence of communities and behavioural change modeled with activity reduction. We consider both the case in which adoption is assigned randomly (red) and in increasing order of activity (blue). Vertical dashed lines indicate the analytical threshold in absence of communities computed from Eq.~\ref{thre3}. We set $\gamma=0.8$, $\psi=0.1$, $p=w=0.6$, $\eta=0.6$, $s=10$. b) Epidemic size as a function of $\lambda$ in presence of communities and behavioural change modeled with modularity increase. We set $\eta=0.6$, $\eta_{\gamma}=0.68$, $\eta_{\psi}=0.96$, $p=w=0.6$. c) Ratio between the final epidemic size obtained in the MI and the AR case as a function of $\eta_{\gamma}$. We set $\lambda=0.6$, $\eta=0.6$, $\eta_{\psi}=0.99$, $p=w=0.6$. In all figures, results are obtained from $10^{2}$ stochastic simulations for each point and with the following common model parameters $\mu=10^{-2}$, $\epsilon=10^{-3}$, $\alpha=2.1$, $N=10^{5}$ and $i_0=I_0/N = 0.01$ (initial fraction of infected seeds).}
  \label{fig3}
\end{figure}

The explicit membership to communities allows considering also another type of behavioural change where individuals keep the same activity but reduce, as a way to lower the infection risk, ties outside their close circle of friends (community). Therefore, the system moves towards isolated social bubbles. In Fig.~\ref{fig3}-b we show the behaviour of the epidemic size as a function of $\lambda$. All the other parameters are set equal to the previous case. In particular, the default (baseline) value of modularity is $\eta=0.6$. However, the fraction of nodes implementing social distancing measures increases the modularity according to their disease status: susceptible nodes are characterized by $\eta_\gamma>\eta$ and infected nodes by $\eta_\psi>\eta$. In order to compare this scenario with the previous one, we set parameters such that the variation of behaviour of nodes changing behaviour has the same magnitude. More in detail, susceptible nodes change their behaviour by $20\%$ (previously we set $\gamma$ to $0.8$, implying a $20\%$ reduction of activity) and infected by $90\%$. Therefore, behavioural change induce a $\eta_\gamma=0.68$ and $\eta_\psi=0.96$. Also in this case, random adoption has a bigger impact on the spreading.\\

By comparing the y-scales of the two scenarios presented in Fig.~\ref{fig3}-a-b, it is clear that the reduction of activity (AR) implies a lower epidemic size with respect to the increase of modularity (MI). We further investigate this point in Fig.~\ref{fig3}-c. To compare these two very different kinds of behavioural change, we fix all the parameters and we study the ratio of the epidemic sizes as a function of modularity of susceptible nodes. In particular, we set $\lambda=0.6$, $\eta=0.6$, $\eta_\psi=0.99$, $p=w=0.6$, and plot $r_\infty^{MI}/r_\infty^{AR}$ as a function of $\eta_\gamma$. In other words, we compare the epidemic size obtained when $60\%$ of nodes reduce activity with what happens when they increase network modularity. The first observation is that, both in case of random and activity-based adoption, the MI strategy leads to higher epidemic sizes. Indeed, the obtained ratio $r_\infty^{MI}/r_\infty^{AR}$ is always greater than one. However, while in the case of random adoption increasing the modularity of susceptible nodes has a strong effect on the epidemic size (which progressively decreases), this effect is not observed in case of activity-based adoption. This result further confirms how the spreading patterns are largely controlled by nodes in high activity classes that do not comply with the social distancing measures. \\

\section{Conclusions}
\label{conclusion}

In this work, we studied the effects of self-initiated behavioural change on disease resurgence using activity-driven networks as a modeling framework for social interactions. We imagined a population that experienced a first wave of infections that was stopped early through strict top-down interventions and did not develop significant immunity to prevent a second wave. We focused on the reactions of individuals' that, when restrictions are lifted, may adopt behavioural measures aimed at protecting themselves by reducing or changing their social interactions. As we write, this scenario is unfortunately extremely realistic and concerning: after the easing of the strict measures established during spring 2020, most Western countries are now facing the second wave of the COVID-19 pandemic.
\\
More in detail, we modeled behavioural change by reducing the activity of susceptible and infected individuals and, having in mind the idea of social bubbles, by increasing  the share of connections within tight social circles with respect to their baseline. In doing so, we explored the effect of behaviour adoption by considering only a fraction of nodes engaging in behavioural protective measures selected either at random or as function of their activity. In fact, the most socially active nodes, maybe due to the nature of their job, cannot easily modify their behaviours.\\
Considering first the simplest version of activity-driven networks, where links are memory-less and random, we derived the analytical threshold of a Susceptible-Infected-Recovered epidemic model. In doing so, we extended the work done by Rizzo et al~\cite{rizzo2014effect} accounting for non-perfect adoption. Interestingly, in case nodes adapting their behaviours are selected randomly in the population, we found that the expression for the basic reproductive number $R_0$ is symmetrical in a combination of activity reduction and level of adoption of susceptible and infected. Given the high socio-economic cost associated with isolating large numbers of people, this finding underlines the importance of efficient isolation of the infected, especially at the beginning of a possible second wave when their number is relatively small. Furthermore, the numerical simulations showed that, in the settings considered, the final epidemic size was mainly dependent on the reduction in activity of the infected.
In case adherence is assigned in increasing order of activity, we found that an almost perfect level of adoption is needed in order to avoid the disease to rise again. This effect is deeply connected to the high heterogeneity of human interactions, and it highlights that even small levels of lack of adoption by very active nodes may have a huge impact on the spreading. Furthermore, this finding is in line with previous studies of targeted immunization strategies on time-varying networks which show how immunizing the most highly active nodes is extremely effective in hampering the spreading of contagion phenomena~\cite{liu2014controlling,vestergaard2014memory,masuda2017temporal,wang2016statistical}. \\
We then moved to even more realistic scenarios, taking into account the tendency of people to cluster in tightly connected groups. To this aim, we considered a modified formulation of the activity-driven model in which nodes are assigned to communities and tend to establish links inside their community more often than with outside nodes~\cite{nadini2018epidemic}. In this setting, we modeled behavioural change by considering two mechanisms: i) reducing the activity of nodes and ii) keeping the same level of activity but limiting the contacts outside communities and thus increase network's modularity. Using numerical simulations, we observed that the modularity of the network increases the threshold with respect to the previous case. This is in line with past observations on synthetic and real time-varying networks~\cite{nadini2018epidemic,onnela2007structure,karsai2011small,salathe2010dynamics,stegehuis2016epidemic,davis2020phase,han2015epidemic}. Furthermore, random adoption is characterized by a larger threshold and a smaller epidemic size with respect to the case of adoption assigned in increasing order of activity. Finally, we found that an activity reduction strategy is more efficient than increasing the modularity across the range of parameters studied. This finding highlights how imperfect social bubbles might not be as effective as an overall reduction of social activities. \\
Of course, this work comes with limitations. First of all, our contribution is only theoretical and limited by the set of assumptions that were made. As such, it should not be indented as a precise representation of reality and especially of the current pandemic landscape. Although we focused on some fundamental features of realistic epidemiological models (e.g. heterogeneity of contacts, modularity, behavioural change), we overlooked many others such as including several connected populations, considering an age-structured population and the complex nature of real self-initiated behavioural change~\cite{hbm1,hbm2,hbm3,funk,verelst,st2020school}. We have limited ourselves to an exploration of the phase space rather than fitting the parameters using real data. Furthermore, we have neglected high-order complex temporal dynamics of real time-varying networks~\cite{lambiotte2019networks,battiston2020networks,petri2018simplicial}. We leave these extensions and model calibration for the future.\\
In conclusion, our work contributes to the characterization of self-initiated behavioural change in the context of disease resurgence on time-varying networks. It highlights the importance of accounting for the heterogeneity of social activation patterns when gauging the efficiency of adaptive strategies aimed at hampering the spreading of infectious diseases on temporal networks.

\section*{Acknowledgements}
All authors thank the High Performance Computing facilities at Greenwich University. N.G. acknowledges support from the Doctoral Training Alliance.

\end{document}